\documentclass[prl,twocolumn,preprintnumbers,amsmath,amssymb,superscriptaddress]{revtex4}

\usepackage{graphicx}
\usepackage{dcolumn} 
\usepackage{bm}      
\usepackage{nicefrac}
\usepackage{color}   

\usepackage{multirow}
\usepackage{longtable}

\def\tc{T_{\rm c}}

\begin{document}

\title{Doped graphane: a prototype high-$\tc$ electron-phonon superconductor}

\author{G. Savini}
\affiliation{Department of Engineering, University of Cambridge, Cambridge, CB3 0FA, UK}
\affiliation{Institute of Protein Research, University of Osaka, Osaka, 565-0871, Japan}

\author{A. C. Ferrari}
\email{acf26@eng.cam.ac.uk}
\affiliation{Department of Engineering, University of Cambridge, Cambridge, CB3 0FA, UK}

\author{F. Giustino}
\affiliation{Department of Materials, University of Oxford, Oxford, OX1 3PH, UK}

\begin{abstract}
We show by first-principles calculations that \emph{p}-doped graphane is a conventional superconductor with a critical temperature ($\tc$) above the boiling point of liquid
nitrogen. The unique strength of the chemical bonds between carbon atoms and the large density of electronic states at the Fermi energy arising from the reduced dimensionality synergetically push $\tc$ above 90K, and give rise to large Kohn anomalies in the optical phonon dispersions. As evidence of graphane was recently reported, and doping of related materials such as graphene, diamond and carbon nanostructures is well established, superconducting graphane may be feasible.
\end{abstract}
\maketitle

The discovery of superconductors such as magnesium diboride\cite{J.Nagamatsu-2001} and iron pnictides\cite{Y.J.Kamihara-2008,H.Takahashi-2008} opened new horizons in the landscape of superconductivity research, fueling renewed interest in the quest for high-temperature superconductivity in materials other than the copper oxides\cite{moussa06,lee06}. The critical temperature, $\tc$, reflects the energy scale of the quantum-mechanical interactions driving the electron condensation into the superconducting state\cite{G.Tinkham-1996}. In high-$\tc$ copper oxides\cite{bednorz86} the nature of the interaction leading to superconductivity is still under debate\cite{F.Giustino-2008}, yet it is generally accepted that Coulomb exchange and correlation effects, with energy scales around few hundred meVs, play an important role\cite{P.W.Anderson-2007,T.A.Maier-2008}. In contrast, in conventional superconductors the pairing is known to be driven by the interaction between electrons and lattice vibrations, with an associated energy scale of only a few ten meVs\cite{J.Bardeen-1957}. Due to the order-of-magnitude difference between such energy scales, it is generally assumed that conventional superconductors cannot exhibit $\tc$ as high as copper oxides\cite{moussa06,lee06}. Here, we report first-principles calculations showing that $p$-doped graphane would make a conventional superconductor with $T_c$ well above the boiling point of liquid nitrogen.

The Bardeen-Cooper-Schrieffer (BCS) theory\cite{J.Bardeen-1957} defines the basic theoretical framework to understand conventional superconductivity. Its generalization, known as the Migdal-Eliashberg theory\cite{P-B.Allen-1982},
incorporating the lattice dynamics, provides a predictive computational tool. Within BCS, $\tc$ is given by\cite{J.Bardeen-1957}:
\begin{equation}\label{Tc}
k_{B}\tc=1.14\hbar \omega_{0}\exp\left(-\frac{1}{N_{F}V}\right)
\end{equation}
where $k_{\rm B}$ is the Boltzmann constant, $\hbar\omega_{0}$ a characteristic phonon energy, $N_{\rm F}$ the electronic density of states (EDOS) at the Fermi Energy, $E_F$, $V$ an effective pairing potential resulting from the net balance between the attractive electron-phonon coupling (EPC) and the repulsive electron-electron interaction\cite{J.Bardeen-1957}. Even though the original BCS formula for $\tc$ is now replaced by more refined expressions such as, e.g., the modified McMillan equation\cite{P.B.Allen-1975}, Eq.\ref{Tc} still proves useful for discussing trends. Eq.\ref{Tc} indicates that one could maximize $\tc$ by increasing the materials parameters $\omega_{0}$, $N_{\rm F}$, $V$. However, these are strongly intertwined, making such optimization complex\cite{P.B.Allen-1975,pickett06}. Here, we propose a simple procedure, based on Eq.\ref{Tc}, to design a high-$\tc$ superconductor.
\begin{figure}[t]
\centerline{\includegraphics[width=80mm]{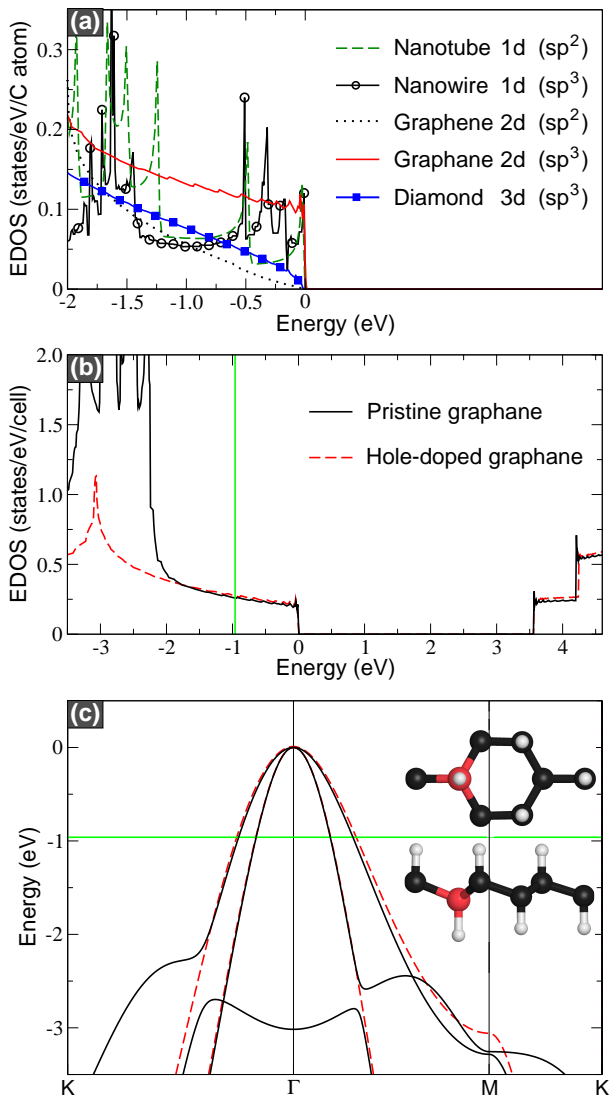}}
\caption{\label{figure1}(Color online)(a)EDOS per carbon atom of 1d (nanotube; diamond nanowire),2d (graphene; graphane) and 3d (diamond) systems. With the exception of graphene, with linear dispersions, the EDOS is proportional to $E^{-\nicefrac{1}{2}}$ in 1d, a step-like function in 2d, and $E^{\nicefrac{1}{2}}$ in 3d. The step-like EDOS in graphane implies that $N_{\rm F}$ is large even at low doping.(b)EDOS of pristine (solid black line) and 12.5\% p-doped graphane (dashed red line). The top of the valence band is set as zero, and $E_F$=-0.96eV (green line). The EDOS at $E_F$ is similar in the two models (0.26 states/eV/cell in rigid-band and 0.27 states/eV/cell in supercell).(c)Band structure of pristine (solid black line) and 12.5\% p-doped graphane (dashed red line). (inset) Ball-and-stick $2\times2$ supercell with one substitutional B(top and side views)}
\end{figure}

Let us first consider the conventional superconductor with the highest $\tc$, MgB$_2$ ($\tc=39$K)\cite{J.Nagamatsu-2001}. For simplicity, we neglect multi-band and anisotropy effects, which were the object of detailed investigations\cite{H.J.Choi-2002,J.M.An-2001,J.Kortus-2001,K.-P.Bohnen-2001,A.Floris-2005}. In MgB$_2$ the EPC contribution to $V$ is large ($\sim$1.4eV, from $\lambda=N_FV$, using $N_F=0.7$states/cell/eV, and $\lambda\sim1$\cite{J.M.An-2001}) because the states with energy close to $E_F$ (those which condensate in the superconducting state\cite{J.Bardeen-1957}) are of $\sigma$ character, i.e. derive from bonding combinations of planar B $sp^{2}$ hybrids localised around the middle of B-B bonds\cite{H.J.Choi-2002,J.M.An-2001,J.Kortus-2001,K.-P.Bohnen-2001,A.Floris-2005}. These electronic states are significantly affected by the B-B bond length variation associated with bond-stretching $E_{2g}$ phonons\cite{yildirim01,J.M.An-2001}, resulting into a large EPC contribution to $V$. At the same time, the $E_{2g}$ phonon energy is large ($\sim60$meV\cite{J.M.An-2001}), due to the small B mass, leading to a large $\omega_{0}$ in Eq.\ref{Tc}. Furthermore, MgB$_2$ is a metal with a significant EDOS at $E_F$($\sim$0.7states/cell/eV\cite{J.M.An-2001}). These three factors cooperate in Eq.\ref{Tc} to establish a superconducting state with $\tc=39$K\cite{H.J.Choi-2002,J.M.An-2001,J.Kortus-2001,K.-P.Bohnen-2001,A.Floris-2005}. However, many attempts to improve upon MgB$_2$, by investigating related materials, only met limited success\cite{H.J.Choi-2009}, with the experimental $\tc$ never higher than MgB$_2$.

We thus search for an alternative material having at least some of the desirable features of MgB$_2$, i.e.(i)$\sigma$ electrons at the Fermi surface,(ii)large bond-stretching phonon frequencies, and (iii)large EDOS at $E_F$. We note that the first two requirements are both met by B-doped diamond, a conventional BCS superconductor with $\tc=4$K\cite{E.A.Ekimov-2004}, where a small hole-like Fermi surface appears around the top of the valence band\cite{K.-W.Lee-2004}. The electronic states at $E_F$ have $\sigma$ character deriving from the bonding combination of tetrahedral C $sp^{3}$ hybrids, bearing some analogy to MgB$_2$. As these $\sigma$ states
are localized in the middle of the C-C bonds, they couple considerably to bond-stretching phonons\cite{K.-W.Lee-2004,L.Boeri-2004}, resulting in a large EPC contribution to $V$, even superior to MgB$_2$ ($\sim3$eV, from $\lambda=N_FV$, using $N_F=0.1$states/cell/eV, and $\lambda\sim0.3$\cite{F.Giustino-2007b})\cite{mauri08,blase09}. In addition, the light C atoms have high energy optical phonons ($\sim130$meV, even after softening induced by the large EPC\cite{F.Giustino-2007b,L.Boeri-2004}). However in B-doped diamond the EDOS at $E_F$ is rather small ($\sim$0.1states/cell/eV for 2\% doping\cite{K.-W.Lee-2004, X.Blase-2004}). This compromises $\tc$. Thus, while B-doped diamond shares some of the desirable features of MgB$_2$, its 3-dimensional (3d) nature implies that the EDOS in proximity of the valence band scales as$\sim E^{\nicefrac{1}{2}}$ (with $E$ measured from the valence band edge)\cite{sutton93}, Fig.\ref{figure1}(a). Then, the number of carriers available for the superconducting state remains relatively small even for large doping. Superconducting diamond is thus a 3d analogue of MgB$_2$\cite{K.-W.Lee-2004,L.Boeri-2004}.

This leads to the question of what would happen in a hypothetical B-doped diamond structure with reduced dimensionality, such as a thin film or a nanowire, where the EDOS can be significantly enhanced by quantum confinement. Indeed, the EDOS of a two-dimensional semiconductor goes as$\sim\theta(E)$ ($\theta$ being the step function)\cite{sutton93}, hence the number of available carriers can be large, even at low doping. In order to estimate the expected EDOS increase in a diamond thin film it is helpful to consider a simple parabolic band model. For 2\% B doping, bulk diamond has $N_F$=0.1 states/eV/cell at $E_F$. A 0.5nm thick diamond film with the same doping would have $N_F\sim$0.5 states/eV/cell. Such an EDOS increase would significantly enhance $\tc$. Using the electron-phonon potential and the phonon frequency of bulk diamond, Eq.\ref{Tc} gives that a 0.5nm film would superconduct at $\tc\sim80$K. However the question remains whether it is possible to synthesize an atomically thin diamond film.

Recent work on graphene and its derivatives points to a positive answer. Soon after the discovery of graphene\cite{novopnas} several works considered how to functionalise and chemically modify this novel 2d material\cite{Geimrev,dai,cervantes,brus,kern,das}. In particular, it was proposed that fully hydrogenated graphene (graphane) would be stable\cite{J.O.Sofo-2007}. The main difference between graphene and graphane is that, while the former is fully $sp^{2}$ bonded, the latter is $sp^{3}$, as diamond\cite{J.O.Sofo-2007}. Recently, some experimental evidence of graphane was reported\cite{D.C.Elias-2009}. Since graphane is the 2d counterpart of diamond, our scaling arguments immediately point to doped graphane as a potential high-$\tc$ superconductor. Doping could be achieved by gating, including using an electrolyte gate, or by charge-transfer, as done in graphene\cite{Geimrev,das,brus,wehling,lanzara,kern}. Substitutional doping of graphene was also reported, up to $\sim10^{14}$ cm$^{-2}$\cite{dai,dai2}.

We thus perform density functional perturbation theory (DFPT) calculations of EPC and superconductivity in doped graphane within the framework of the Migdal-Eliashberg theory\cite{P-B.Allen-1982} and the local density approximation (LDA)\cite{D.M. Ceperley-1980,A.Zunger-1980}. By analogy with B doped diamond, we consider $p$-doping. This is simulated using the rigid-band approximation\cite{J.Noffsinger-2009}. Fig.\ref{figure1}(b) shows that the calculated EDOS in \emph{p}-doped graphane close to the valence band maximum follows a step-like behavior, as expected for a 2d system. At 3\% doping the EDOS is 0.22 states/eV/cell, compared to 0.13 states/eV/cell in bulk diamond, with a factor 1.7 enhancement. Fig.\ref{figure1}(c) indicates that the dispersions close to $E_F$ are essentially identical for a supercell containing B and for a rigid-band model of doped graphane. We expect this to hold also for lower doping, where the perturbation to the pristine dispersions is smaller. The similarity between these two models justifies our use of the rigid-band approximation. A supercell calculation with the B dopant explicitly included does not show impurity states inside the gap.
\begin{figure}[t]
\centerline{\includegraphics[width=90mm]{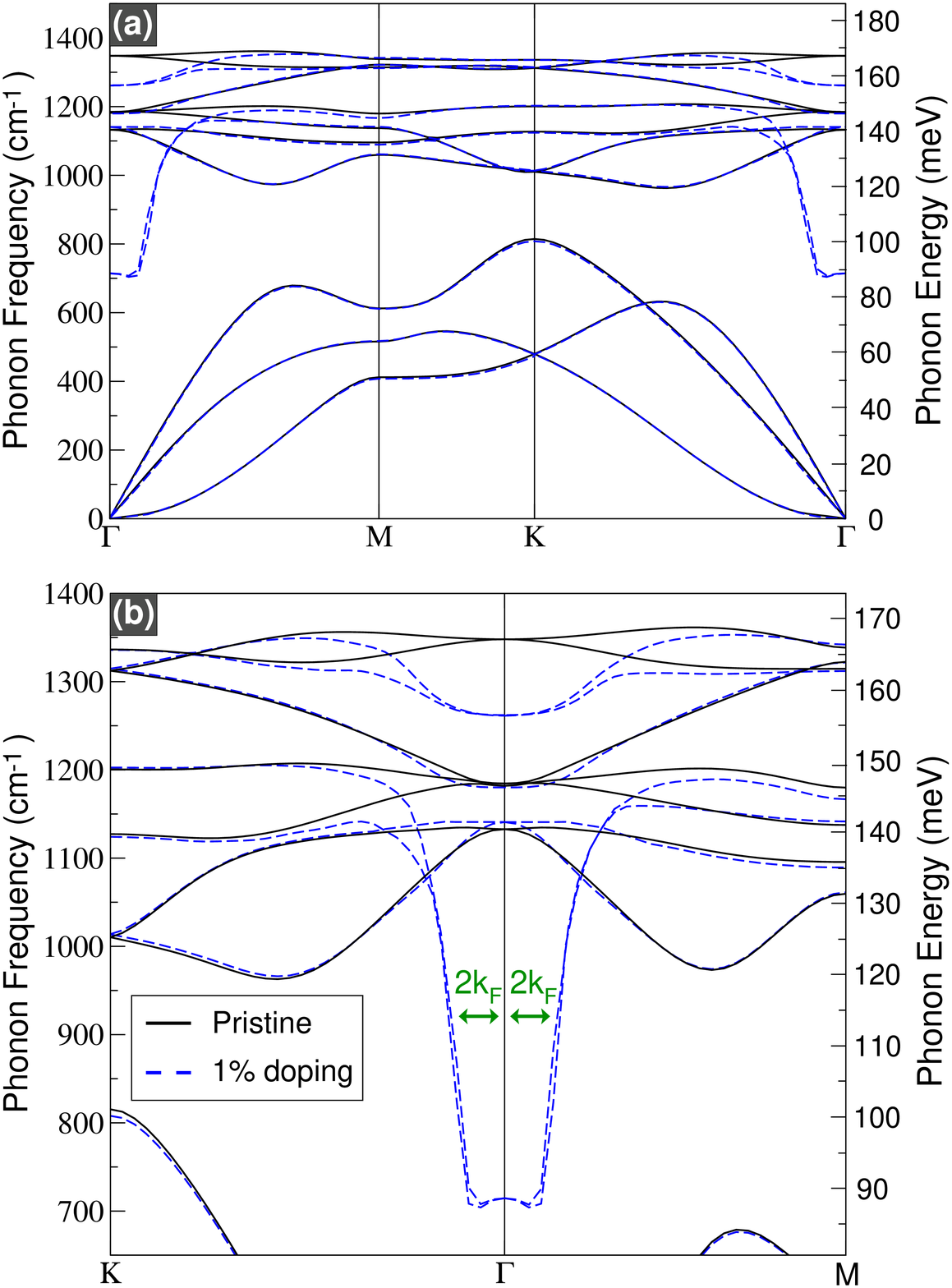}}
\caption{\label{figure2a}(Color online)(a)Phonon dispersion of pristine (solid black line) and 1\% $p$-doped graphane (dashed blue lines). The C-H stretching modes have higher frequencies (2655-2711cm$^{-1}$) and are not shown.(b)Optical modes around the zone centre, showing the Kohn Anomalies. The horizontal (green) arrows indicate the average Fermi surface diameter.}
\end{figure}
\begin{figure}[t]
\centerline{\includegraphics[width=85mm]{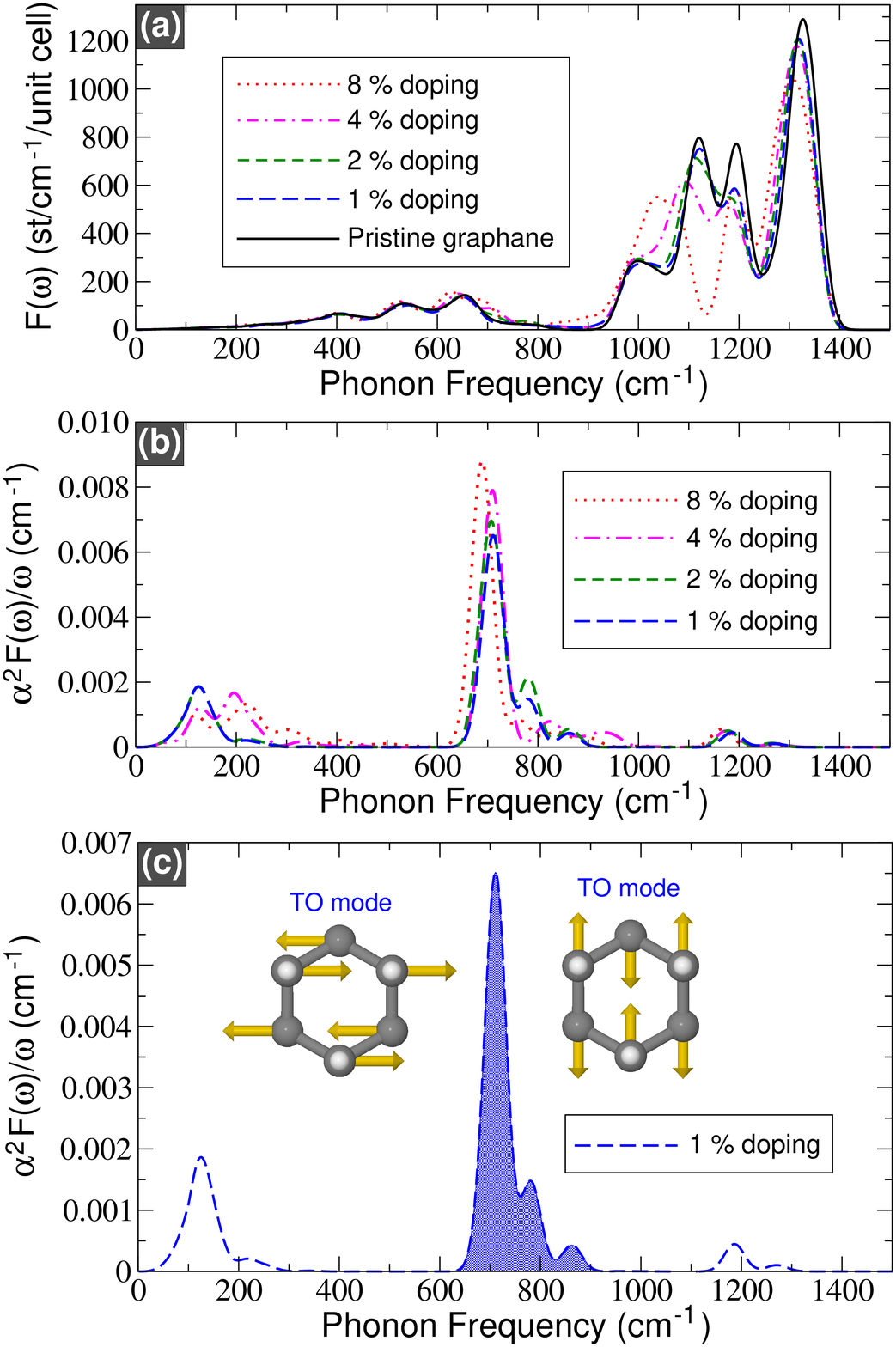}}
\caption{\label{figure2b}(Color online)(a)PDOS of pristine and doped graphane.(b) Eliashberg function in $p$-doped graphane for increasing doping. The largest contribution comes from the optical modes, similar to diamond\cite{L.Boeri-2004}, but also the acoustic phonons couple to holes at the Fermi surface, similar to SiC\cite{J.Noffsinger-2009}.(c)Contributions to the Eliashberg function arising from the TO stretching modes (hashed region). (insets)  Ball-and-stick representations of two TO modes. The arrows indicate the in-plane C-C stretching motions (carbons are shown in grey, hydrogens in white)}
\end{figure}

Fig.\ref{figure2a}(a,b) report the phonon dispersions of pristine and $p$-doped graphane and Fig.\ref{figure2b}(a) the corresponding phonon density of states (PDOS). Upon doping, the optical zone-centre modes with in-plane C-C stretching soften as a result of the inception of Kohn Anomalies\cite{W.Kohn-1959}. The two degenerate TO modes, having planar C-C stretching and H atoms moving in-phase with the C atoms, downshift from 1185 to 715cm$^{-1}$ (147 to 89meV). This is due to the large EPC of planar C-C stretching, which significantly affects the $sp^3$-like electronic states at the Fermi surface. The two degenerate zone-centre modes, having in-plane C-C stretching and H atoms moving out-of-phase with respect to the C atoms, downshift from 1348 to 1257cm$^{-1}$ (167 to 156meV). The LO mode, with out-of-plane C-C stretching, does not couple to the electrons due to the different parity of potential and wavefunctions, resulting into a vanishing EPC. The two degenerate optical modes corresponding to the shear motion of the C and H planes (at$\sim$1133cm$^{-1}$) and the C-H stretching modes (2 modes at 2663 and 2711cm$^{-1}$) do not undergo softening upon doping. This is consistent with the electronic states associated with the C-H bonds having little weight at $E_F$, hence a small EPC.
\begin{figure}[th]
\centerline{\includegraphics[width=80mm]{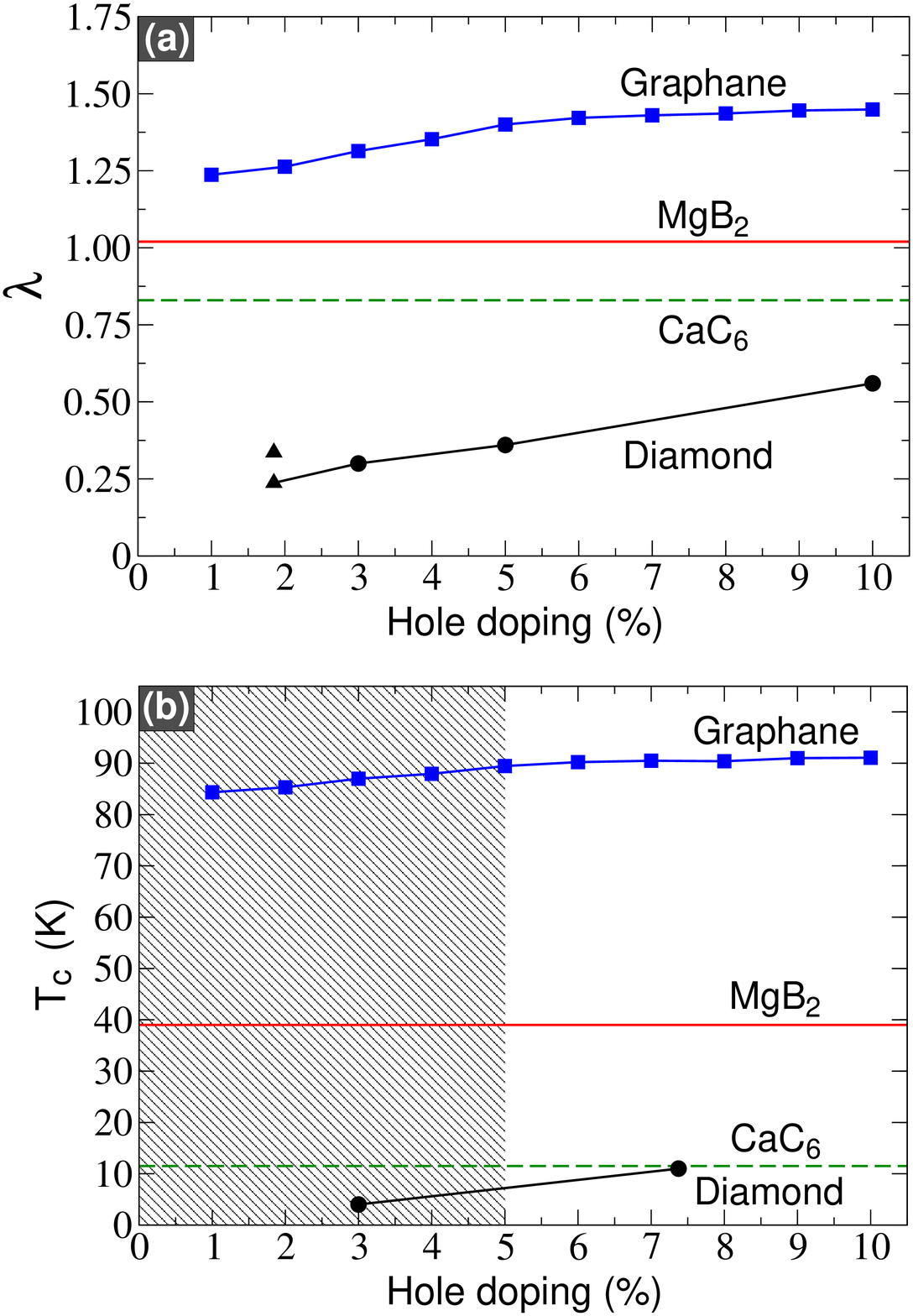}}
\caption{\label{figure3}(Color online)(a)EPC of graphane as a function of doping, calculated using the standard DFPT formalism\cite{baroni03}: the Brillouin zone is sampled with an electron grid up to $300\times300\times1$, smearing from 50 to 270 meV, and phonon grid of $100\times100\times1$. For comparison, we plot literature values for MgB$_2$ (solid red line\cite{L.Boeri-2004}), CaC$_6$ (dashed green line\cite{M.Calandra-2006}), and diamond (solid black line\cite{L.Boeri-2004}; triangles\cite{F.Giustino-2007b}). More sophisticated calculations taking explicitly into account a substitutional dopant, such as B, could slightly change the EPCs\cite{X.Blase-2004,F.Giustino-2007b}. However, in B-doped diamond a rigid-band model provides a lower EPC and a lower bound to $\tc$\cite{F.Giustino-2007b}.(b)$\tc$ calculated using the modified McMillan formula and a Coulomb pseudopotential $\mu{^*}=0.13$\cite{G.Grimvall-1981}. The left-side hashed region indicates doping below the estimated MIT, where our formalism applies only to charge-transfer or gate-induced doping. Above MIT, it applies to substitutional doping as well. We use the isotropic Eliashberg formalism\cite{G.Grimvall-1981}. A more sophisticated description based on the fully anisotropic Eliashberg theory is expected to increase $\tc$\cite{H.J.Choi-2002,choi-prb}. For comparison we also show $\tc$ of MgB$_2$ (solid red line, $\tc=39$K\cite{J.Nagamatsu-2001}), CaC$_6$ (dashed green line, $\tc=11.5$K\cite{N.Emery-2005}), and diamond (solid black line, $\tc=4$K at$\sim$3\% B\cite{E.A.Ekimov-2004}; 11K at$\sim7$\% B\cite{K.Ishizaka-2007}}
\end{figure}

The softening of modes with a large C-C stretching component is similar to that reported in B-doped diamond\cite{L.Boeri-2004,F.Giustino-2007b}. In particular, the region of reciprocal space where the phonon softening is observed matches the diameter, $2k_F$, of the hole Fermi surface around the $\Gamma$ point, this being a typical signature of the Kohn effect\cite{W.Kohn-1959}. The calculated phonon softening of the TO C-C stretching modes ($\sim58$meV or$\sim470$cm$^{-1}$) is significantly larger than in other materials, as typical Kohn anomalies range from $\sim5$meV (graphite and graphene\cite{S.Piscanec-2004}) to$\sim 10$meV (TaC \cite{J.Noffsinger-2008}). In the case of B-doped diamond the phonon softening takes place through the creation of a non-dispersive defect branch associated with the B dopant\cite{F.Giustino-2007b}. A similar effect could happen in B-doped graphane, but we expect the magnitude of the doping-induced softening to be reasonably well described within our rigid-band model. Also, more sophisticated calculations, taking B explicitly into account\cite{X.Blase-2004,F.Giustino-2007b} or with non-adiabatic corrections\cite{saitta08}, may slightly revise the softening. Nevertheless, such a large softening stands out as a qualitative effect.

Figure \ref{figure2b}(b) plots the Eliashberg spectral function\cite{G.Grimvall-1981}, which shows the relative contribution of different modes to the superconducting pairing\cite{G.Grimvall-1981}:
\begin{equation}
\alpha^{2}F\left(\omega\right)=\dfrac{1}{2}\underset{\mathbf{q}\nu}{\sum}\omega_{\mathbf{q}\nu}\lambda_{\mathbf{q}\nu}\delta\left(\omega-\omega_{\mathbf{q}\upsilon}\right)
\end{equation}
where $\lambda_{\mathbf{q}\nu}$ is the EPC for a phonon mode $\nu$ with momentum $\mathbf{q}$ and frequency $\omega_{\mathbf{q}\nu}$, and
$\delta$ is the Dirac delta (we used a Gaussian of width 2meV for this purpose). We get that the TO in-plane C-C bond-stretching phonons with C and H atoms moving in-phase (see Fig.\ref{figure2b}(c)) have the largest EPC, due to the $\sigma$ character of the electronic states at $E_F$ and the large C displacements associated with these modes. This is similar to B-doped bulk diamond\cite{K.-W.Lee-2004,L.Boeri-2004,X.Blase-2004,F.Giustino-2007b} and validates our hypothesis that $p$-doped graphane can be regarded as an atomically thin diamond film, exhibiting similar EPC and vibrational frequencies, but larger EDOS at $E_F$. We note that the in-plane C-C bond-stretching phonons, with C and H atoms moving out-of-phase, do not contribute to the EPC. This happens because, upon softening, the four C-C planar stretching modes hybridize in such a way that those at 715cm$^{-1}$ carry an increased weight on the C atoms, while the opposite happens for the two modes at 1257cm$^{-1}$.

Figure \ref{figure3}(a) plots the EPC as a function of doping, and Fig. \ref{figure3}(b) the corresponding $\tc$. We find that $\tc$ exceeds the boiling point of liquid nitrogen, and falls within the same $\tc$ range as copper oxides\cite{pickett89}. Due to the relatively constant EDOS below the top of the valence band [cf. Fig.\ref{figure1}(b)], $\tc$ is rather insensitive to doping. This is important for the practical realization of superconducting graphane. Our results should be valid throughout the entire doping range considered here in the case of gate- or charge transfer-induced doping, since in these cases the holes are delocalized and doped graphane is in the metallic regime. On the other hand, for substitutional doping we expect our results to be valid only beyond the Mott metal-to-insulator transition (MIT). In absence of experimental MIT measurements in graphane, we estimate the critical doping concentration, $n_c$, using the following argument. In 3d the MIT occurs when the impurity wavefunctions are close enough that their overlap is significant\cite{bustarret08}. For many materials $a_Hn_c^{1/3}\sim0.26$, $a_H$ being the radius of the ground-state wavefunction of an hydrogenic donor\cite{bustarret08}. The radius can be calculated as $a_H=\epsilon/m^\star a_0/2$, $a_0$ being the Bohr radius, $\epsilon$ the dielectric constant, and $m^\star$ the effective mass\cite{bustarret08}. In diamond $a_H\sim4$\AA\ and $n_c\sim4\cdot20$cm$^{-3}$\cite{bustarret08}, therefore the average separation between nearest neighbor B atoms is$\sim15$\AA. In the case of graphane we use a similar criterion, replacing the 3d hydrogenic impurity with a 2d one. The ground-state hydrogenic wavefuction in 2d has a radius $a_H^{2d}=\epsilon/m^\star a_0/2$\cite{yang91}. Using the dielectric constant and hole effective mass of diamond ($\epsilon=5.7$; $m^\star=0.74$) we find $a_H^{2d}=a_{H}/2\sim2$\AA. Thus, the average separation between nearest neighbor B atoms at the MIT is $\sim7.5$\AA\, and the corresponding doping can be estimated as 5\% B (1 B every 20 C atoms) or $2\cdot10^{14}$ holes$\cdot$cm$^{-2}$. This could be feasible, considering that substitutional doping in graphene was reported up to 5\%\cite{dai2}.

The calculated high-$\tc$ for \emph{p}-doped graphane bears consequences both for fundamental science and applications. One could envision hybrid superconducting-semiconducting circuits directly patterned through lithographic techniques, graphane-based Josephson junctions for nanoscale magnetic sensing, and ultimately an ideal workbench for exploring the physics of the superconducting state in two dimensions\cite{S.Qin-2009}. The superconducting phase transition in graphane could also be controlled by gating\cite{iwasa09,das}. A high-$\tc$ superconductor with gate-controllable $\tc$ could lead to novel switching mechanisms in nanoscale field-effect transistors. Furthermore, the discovery of an electron-phonon superconductor with $\tc$ above liquid nitrogen would mean that(i) there are no fundamental reasons to believe that BCS superconductors cannot have $\tc>$40K (MgB$_2$), and (ii)high-$\tc$ superconductivity does not take place exclusively in the copper oxides. In particular, our calculations indicate that at least one material could exist where a very strong EPC leads to $\tc$ in the copper oxide range without triggering a lattice instability. The superconducting phase transition in systems with reduced dimensionality has been the subject of numerous theoretical studies\cite{rice65,hohenberg66}. Quantum fluctuations could destroy the superconducting order in 2d\cite{tinkham75}. However, recent experimental evidence suggests that this is not necessarily the case\cite{S.Qin-2009,iwasa09,bozovic08}. In particular, for thin Pb it was reported that the superconducting state is robust down to two atomic layers\cite{S.Qin-2009}. Since our proposed mechanism of superconductivity in doped graphane is BCS-like, as in Pb, there should be no fundamental limits to prevent the realization of high-$\tc$ superconductivity in graphane.

It is immediate to extend the present study to diamond nanowires, which have been the subject of intense investigations in the past few years\cite{yang08}. For a 1d system the EDOS near a band edge has a van Hove singularity going as$\sim E^{-\nicefrac{1}{2}}$\cite{sutton93}. We can assume phonon energies and EPC to be similar to
bulk diamond and graphane. Then Eq.\ref{Tc} would yield $\tc$ as high as$\sim150$K for a 1nm nanowire (see EDOS in Fig.\ref{figure1}(a)).
The possibility of achieving $\tc$ higher than copper oxides by exploiting dimensionality deserves further investigation. Our work suggests that \emph{p}-doped diamond nanostructures have an intriguing potential for high-$\tc$ BCS-like superconductivity.

\textbf{Acknowledgments}
Calculations were performed at HPCF (Cambridge) and the Research Center for Scientific Simulations of the University of Ioannina (Greece) using {\tt Quantum ESPRESSO}\cite{qe09}. ACF acknowledges funding from The Royal Society, the EU grant NANOPOTS and EPSRC grant EP/G042357/1; GS from JSPS and Grant-in-Aid for Scientific Research.

\end{document}